\documentclass[11pt]{article}
\usepackage{moriond,epsfig}

\def\be{\begin{equation}}
\def\ee{\end{equation}}
\def\bea{\begin{eqnarray}}
\def\eea{\end{eqnarray}}

\begin{document}
\vspace*{4cm}

\title{TWO-BODY NONLEPTONIC B DECAYS IN THE STANDARD MODEL AND BEYOND}

\author{M. CIUCHINI${\,}^1$, E. FRANCO${\,}^2$, G. MARTINELLI${\,}^2$,
  A. MASIERO${\,}^3$, M. PIERINI${\,}^2$, L. SILVESTRINI${\,}^2\,$\footnote{Talk
  given by L. Silvestrini}}

\address{ 
${}^1\!\!$ {\em INFN
    Sezione di Roma III and Dip. di Fisica, Univ. di Roma Tre,}\\
{\em Via della Vasca Navale 84, I-00146 Rome, Italy.}\\
${}^2\!\!$
{\em INFN, Sez. di Roma and Dip. di Fisica, Univ. di Roma ``La Sapienza''} \\
{\em Piazzale Aldo Moro 2, 00185 Rome, Italy. } \\
${}^3\!\!$ {\em Dip. di Fisica ``G.
    Galilei'', Univ. di Padova and INFN,}\\ 
  {\em Sezione di
    Padova, Via Marzolo 8, I-35121 Padua, Italy.}}

\maketitle\abstracts{ We briefly discuss the phenomenology of $B \to
  \pi \pi$, $B \to K \pi$ and $B \to \phi K$ decays in the Standard
  Model and in Supersymmetry.}

\section{Introduction}
\label{sec:intro}

After a few years of very successful running of $B$-factories, and
with the bright prospects of experiments at the Tevatron, at the LHC
and, hopefully, at a super $B$-factory, $B$ physics is playing a
central role in testing the Standard Model (SM) and looking for new
physics. In addition to the leptonic and semileptonic modes, it is
certainly useful to use nonleptonic decays for this purpose, given the
large amount of experimental data available on a huge variety of
channels. Nonleptonic decays however pose serious theoretical
challenges, since one must get rid of all the hadronic uncertainties
due to the presence of exclusive hadronic final states in order to
extract information on short-distance dynamics. Indeed, apart for a
handful of golden channels in which hadronic uncertainties drop in CP
asymmetries, such as the celebrated $B \to J/\psi K_s$ and, within the
SM, $B \to \phi K_s$, we have to face the difficulty of estimating
hadronic matrix elements including final state interactions, which in
particular may play a crucial role in CP asymmetries.

These considerations have stimulated an intense theoretical activity
in the last few years, leading to various approaches to the
computation of two-body nonleptonic $B$ decays: QCD
factorization,~\cite{BBNSpipi,BBNS} pQCD,~\cite{li} Soft-Collinear
Effective Theory (SCET).~\cite{SCET,pipiscet} Based on different
assumptions on the relevant degrees of freedom and on the
calculability of certain matrix elements, all these approaches have
lead to factorization theorems for $B$ decays to two light mesons in
the limit of infinite $B$-quark mass. These beautiful theoretical
results can be safely and succesfully applied to compute from first
principles decay amplitudes in which power-suppressed terms are not
accidentally enhanced and thus remain at the level of $10-20 \%$. On
the other hand, most of the phenomenologically interesting channels
contain penguin amplitudes, and in particular penguin contractions of
current-current operators containing charm quarks (charming penguins).
While no rigorous treatment of these contributions has been given in
pQCD until now, in the QCD factorization approach these penguins are
considered to be perturbatively calculable up to power suppressed
terms, while it has been recently pointed out that charming penguins
arise as leading-order nonfactorizable contributions in
SCET.~\cite{pipiscet} Adding this to the fact that charming penguins
are doubly Cabibbo enhanced in $b \to s $ decays, it is clear that, as
it was pointed out long ago,~\cite{charmingpipi,charming} these
processes are most probably dominated by non-calculable charming
penguins, and even $b \to d$ transitions, where this Cabibbo
enhancement is absent, might be affected by large theoretical
uncertainties.

These observations can be tested with the help of experimental data:
one can implement, for example, QCD factorization formulae for $B$
decays to two light mesons, add to these a parameterization of
dominant nonfactorizable amplitudes, and check if this gives a
satisfatory description of experimental data. If this is the case, it
is also possible to quantify the size of nonfactorizable terms and to
test the consistency of the factorization theory. One can also,
varying these nonperturbative terms in a reasonable range, compute
$B$ decays in models beyond the SM, for example in Supersymmetry
(SUSY), and quantify the deviations from the SM prediction in
particularly sensitive quantities, as for example the CP asymmetry in
$B \to \phi K_s$. In this talk, we will illustrate these points with
some significant examples: $B \to \pi \pi$ decays, $B \to K \pi$
decays and ${\cal A}_{CP}(B \to \phi K_s)$. A more general and
comprehensive analysis can be found in ref.~\cite{inprogress}

\section{$B \to \pi \pi$ decays}

$B \to \pi \pi$ decays are particularly interesting since the CP
asymmetry in the $\pi^+ \pi^-$ channel would be proportional to $\sin
2 \alpha$ in the absence of penguins. In the presence of penguins, one
can still use all the available experimental data, which now also
include a measurement of the $BR(B \to \pi^0 \pi^0)$, to identify the
penguin contribution and extract $\alpha$ with some quantifiable
uncertainty. Now, while early QCD factorization studies estimated
$BR(B \to \pi^0 \pi^0)$ to lie in the $10^{-8}-10^{-7}$
range,~\cite{BBNSpipi} it had also been pointed out that
nonperturbative effects could easily bring it up to the level of
$10^{-6}$.~\cite{charmingpipi} In this sense, there is actually no
``$\pi \pi$ puzzle'': adding to the QCD factorization amplitude the
effect of charming and GIM (up minus charm) penguins at the subleading
level, one can perfectly reproduce the observed BR's and asymmetries,
although the present value of BR$(B \to \pi^0 \pi^0)$ is at the upper
end of the expected range. In Table \ref{tab:pipi} we report the
(input) experimental data~\footnote{Concerning the quantity
  $\mathcal{S}_{\pi^+ \pi^-}$, we use a skeptical combination of the
  BaBar and Belle results. All other averages are taken from the
  HFAG.~\cite{HFAG}} and the results of our fit, which corresponds to
charming penguins $P_1 = (0.11 \pm 0.05) e^{i (-0.2 \pm 0.9)}$ and GIM
penguins $P_1^{\mathrm{GIM}} = (0.43 \pm 0.14) e^{i (-0.2 \pm 0.7)}$
in units of the factorized amplitude, using as input the best values
of CKM parameters from the Unitarity Triangle fit
(UTfit~\cite{utfit}). The ``large'' fitted value of
$P_1^{\mathrm{GIM}}$ is not so surprising, taking into account that it
effectively incorporates other nonfactorizable contributions
(annihilations and corrections to emission topologies).~\cite{BS}

\begin{table}[t]
\caption{Fit of $B \to \pi \pi$ observables: BR's, direct CP
  asymmetries ($\mathcal{A}_\mathrm{CP}$) and coefficients of the
  $\sin \Delta M t$ term in time-dependent CP asymmetries ($\mathcal{S}$)}
\label{tab:pipi}
\vspace{0.4cm}
\begin{center}
\begin{tabular}{|c|c|c|c|c|c|c|}
\hline
& & & & & & \\
Channel & $BR^\mathrm{th} \times 10^6$ & $BR^\mathrm{exp} \times 10^6$ &
$\mathcal{A}_\mathrm{CP}^\mathrm{th}$ &
$\mathcal{A}_\mathrm{CP}^\mathrm{exp}$ & $\mathcal{S}^\mathrm{th}$ &
$\mathcal{S}^\mathrm{exp}$\\ 
& & & & & & \\
$\pi^+ \pi^- $& $4.6 \pm 0.4$ & $4.6 \pm 0.4$ & $0.45 \pm 0.13$ &
$0.46 \pm 0.13$ 
& $-0.7 \pm 0.2$ & $-0.7 \pm 0.3$ \\
$\pi^+ \pi^0$ & $5.2 \pm 0.7$ & $5.3 \pm 0.8$ & - & $-0.07 \pm 0.14$ & - & - \\
$\pi^0 \pi^0$ & $1.8 \pm 0.5$ & $1.9 \pm 0.5$ & $-0.27 \pm 0.46$ & - &
- & -\\ 
& & & & & & \\ \hline
\end{tabular}
\end{center}
\end{table}

\begin{figure}
  \begin{center}
    \begin{tabular}{c c}
      \includegraphics[width=0.4\textwidth]{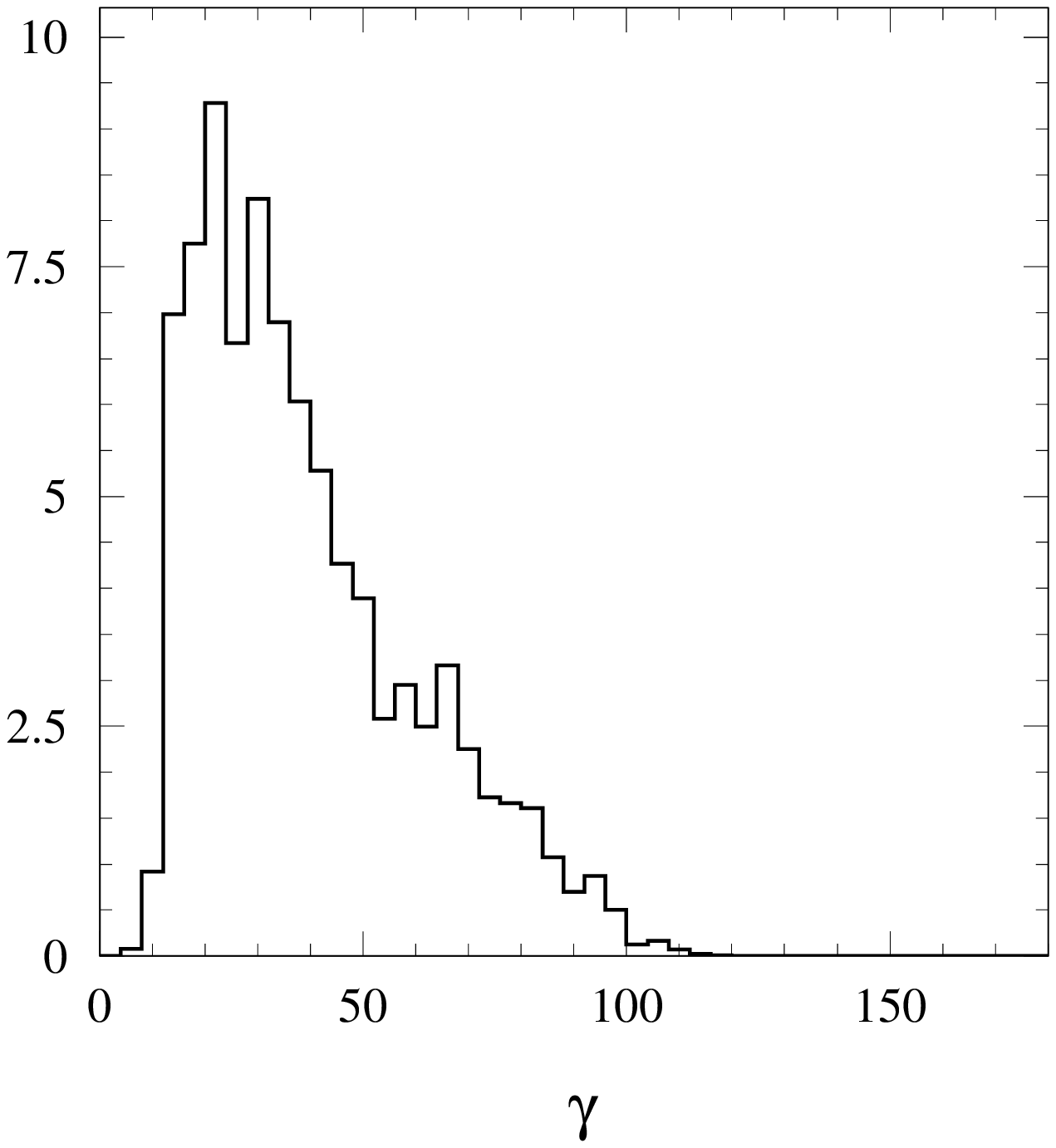} &
      \includegraphics[width=0.4\textwidth]{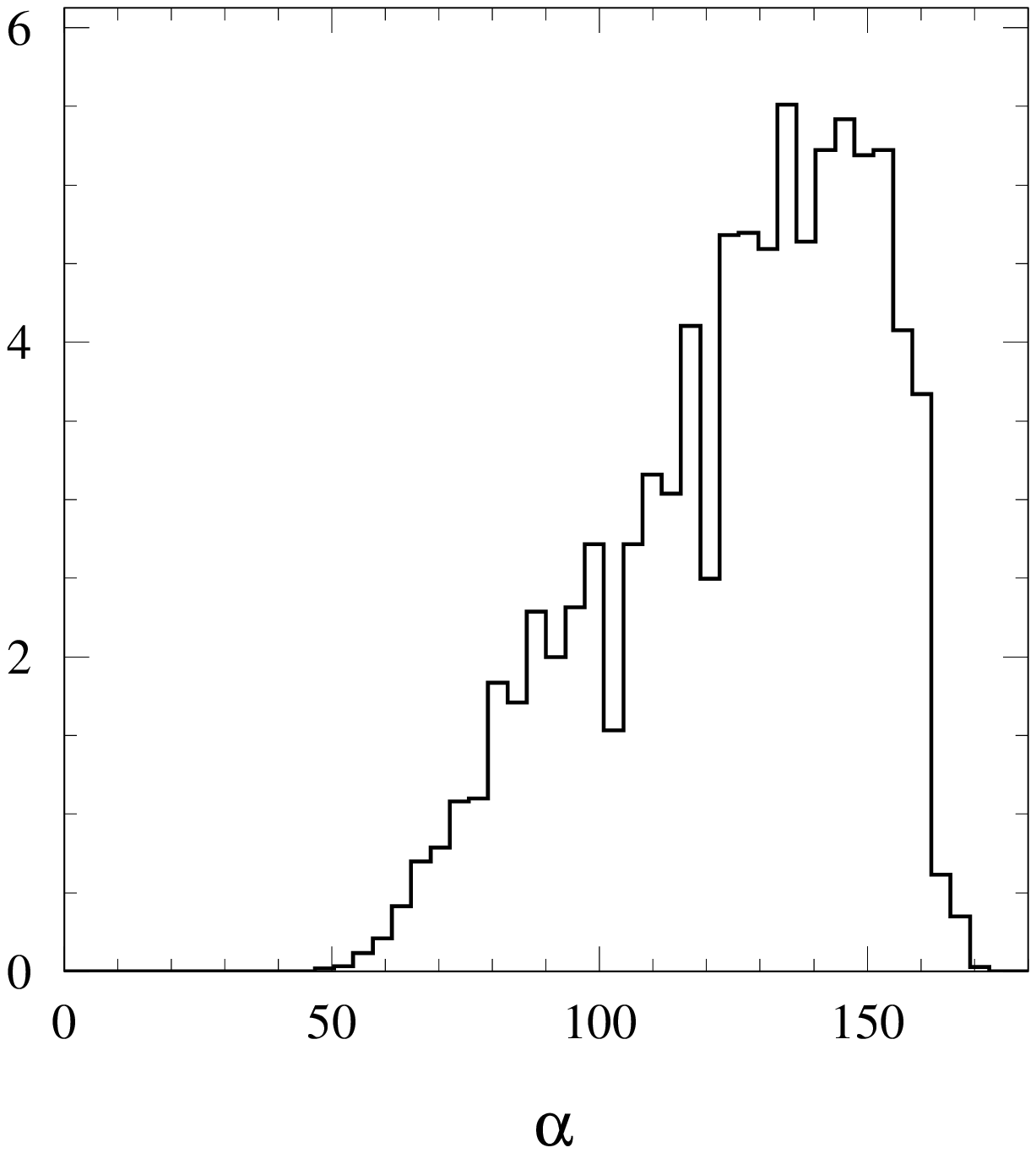} \\
    \end{tabular}
  \end{center}
\caption{Probability density function for $\gamma$ and $\alpha$ as
  extracted from $B \to \pi \pi$ decays, assuming a flat \emph{a
  priori} distribution for $\gamma$.}
\label{fig:gammapipi}
\end{figure}

It is interesting to notice that if one leaves $\gamma$ as a free
parameter in the fit, the {\em a posteriori} distribution for
$\gamma$, or equivalently for $\alpha$, contains some nontrivial
information. In fig. \ref{fig:gammapipi} we report the p.d.f. for the
angles $\gamma$ and $\alpha$. The shape is similar to the one obtained
in a completely model-independent SU(2)
analysis.~\cite{ali,utfitfuture}

\section{$B \to K \pi$ decays}

$B \to K \pi$ decays are particularly interesting as they are
penguin-dominated $b \to s$ transitions in which SM penguin operators
are doubly Cabibbo enhanced with respect to current-current operators.
On one hand, this means that these channels are particularly sensitive
to NP contributions; on the other, as we stressed in the Introduction,
this also implies that nonfactorizable contributions, and in
particular charming penguins, are expected to dominate the amplitude.
This unfortunately introduces large theoretical uncertainties and
possibly spoils the sensitivity to NP. 

Let us now quantify these statements. First of all, we report in Table
\ref{tab:kpi} the theoretical predictions and the experimental values
for $B \to K \pi$ BR's and CP asymmetries. These correspond to a
fitted value of $P_1=(0.08\pm 0.02)e^{i (-0.6 \pm 0.5)}$, while
$P_1^\mathrm{GIM}$ is irrelevant in these channels and therefore not
determined by the fit. 

\begin{table}[t]
\caption{Fit of $B \to K \pi$ observables: BR's and direct CP
  asymmetries ($\mathcal{A}_\mathrm{CP}$).}
\label{tab:kpi}
\vspace{0.4cm}
\begin{center}
\begin{tabular}{|c|c|c|c|c|}
\hline
& & & & \\
Channel & $BR^\mathrm{th} \times 10^6$ & $BR^\mathrm{exp} \times 10^6$ &
$\mathcal{A}_\mathrm{CP}^\mathrm{th}$ &
$\mathcal{A}_\mathrm{CP}^\mathrm{exp}$ \\ 
& & & &\\
$K^+ \pi^- $& $18.7 \pm 0.7$ & $18.2 \pm 0.8$ & $-0.08 \pm 0.03$ & 
$-0.095 \pm 0.028$ \\
$K^+ \pi^0$ & $12.2 \pm 0.6$ & $12.8 \pm 1.1$ & $-0.08 \pm 0.03$ &
$0.00 \pm 0.07$ \\
$K^0 \pi^+$ & $22.2 \pm 0.9$ & $21.8 \pm 1.4$ & $0.00 \pm 0.05$ &
$0.02 \pm 0.06$ \\ 
$K^0 \pi^0$ & $8.7 \pm 0.6$ & $11.9 \pm 1.5$ & $0.03 \pm 0.07$ &
$0.03 \pm 0.37$\\ 
& & & &\\ \hline
\end{tabular}
\end{center}
\end{table}

From Table \ref{tab:kpi}, we see a $\sim 2 \sigma$ deviation in the
$K^0 \pi^0$ channel, which cannot be fixed by any isospin invariant
physics. Indeed, to reproduce the experimental value, an
$\mathcal{O}(1)$ isospin breaking in $P_1$ would be needed.  If
confirmed with increased experimental accuracy, this discrepancy would
call for NP contributions in EW penguins.~\cite{kpipuzzle,npew}

\section{$B \to \phi K$ decays}
\label{sec:kphi}

$B \to \phi K$ decays are pure penguin $b \to s$ transitions. In the
absence of GIM penguins, and neglecting doubly Cabibbo suppressed
terms in $V_{tb} V_{ts}^*$, the decay amplitude in the SM has a vanishing
weak phase and therefore one expects no direct CP violation and the
same time-dependent CP asymmetry as in $B \to J/\psi K$ decays. In our
framework, we can quantify this statement: we add charming and GIM
penguins to the QCD factorization amplitude, fit them to the BR's and
obtain a p.d.f. for the CP asymmetries in $B \to \phi K_s$
decays. Taking as input the UTfit values for CKM angles,~\cite{utfit}
in particular $\sin 2 \beta = 0.710 \pm 0.037$, we obtain the SM
prediction: 
\begin{equation}
  \label{eq:phikssm}
\mathcal{S}_{\phi K_s}=0.73 \pm 0.07\,,\qquad 
\mathcal{C}_{\phi K_s}=0.00 \pm 0.07\,  
\end{equation}
which is fully compatible with the BaBar result 
$\mathcal{S}_{\phi K_s}=0.47 \pm 0.34^{+0.08}_{-0.06}$, but $\sim 3 \sigma$
away from the Belle measurement $\mathcal{S}_{\phi K_s}=-0.96 \pm
0.50^{+0.09}_{-0.11}$.   

The same exercise can be done for $B \to K_s
\pi^0$: fitting the relevant hadronic parameters to $B \to K \pi$ BR's,
we obtain the following SM prediction for the time-dependent
asymmetry, always starting from the UTfit CKM angles:
\begin{equation}
  \label{eq:phikspi}
\mathcal{S}_{K_s \pi^0}=0.79 \pm 0.08\,,\qquad 
\mathcal{C}_{K_s \pi^0}=-0.03 \pm 0.07\,,
\end{equation}
to be compared to the BaBar result $\mathcal{S}_{K_s
  \pi^0}=0.48^{+0.38}_{-0.47} \pm 0.11$.  It should be stressed that,
in the presence of NP, deviations from the SM could differ
considerably in the $K \phi$, $K \pi$ and $K \eta^\prime$ systems.
Indeed, NP contributions can be very sensitive to poorly known
hadronic matrix elements, and large direct CP violation could be
generated in these channels.~\cite{marco}

\section{Beyond the Standard Model: A simple SUSY example}
\label{sec:susy}

If the discrepancy between the experimental value of the
time-dependent asymmetry in $B \to \phi K_s$ and the SM prediction is
confirmed by future measurements at B factories, the question arises
of what kind of NP could account for this deviation. After the
pioneering pre-B factory studies,~\cite{oldphiks} this problem has been
widely studied in the SUSY context in the recent
literature.~\cite{newphiks} Just for the purpose of illustration, we
briefly report the results of a model-independent analysis in the
Minimal Supersymmetric Standard Model (MSSM).~\cite{Ciuchini:2002uv}
Minimality refers here only to the minimal amount of superfields
needed to supersymmetrize the SM and to the presence of R parity.
Otherwise the soft breaking terms are left completely free and
constrained only by phenomenology.  Technically the best way we have
to account for the SUSY FCNC contributions in such a general framework
is via the mass insertion method using the leading gluino exchange
contributions.~\cite{hall} In the Super-CKM basis, SUSY FCNC and CP
violation arise from off-diagonal terms in squark mass matrices only.
These are conveniently expressed as $(\delta_{ij})_{AB}\equiv
(\Delta_{ij})_{AB}/m^2_{\tilde q}$, where $(\Delta_{ij})_{AB}$ is the
mass term connecting squarks of flavour $i$ and $j$ and ``helicities''
$A$ and $B$, and $m_{\tilde q}$ is the average squark mass.

We performed a MonteCarlo analysis, generating weighted random
configurations of input parameters and computing for each
configuration the weight corresponding to the experimental values of
BR$(B \to X_s \gamma)$, $A_{CP}(B \to X_s \gamma)$, BR$(B \to X_s
\ell^+ \ell^-)$ and the $B_s - \bar B_s$ mass difference $\Delta
M_{B_s}$. We study the clustering induced by the contraints on various
observables and parameters, assuming that each unconstrained
$\delta_{23}^d$ fills uniformly a square $(-1\dots 1$, $-1\dots 1)$ in
the complex plane. The ranges of CKM parameters have been taken from
the UTfit, and hadronic parameter ranges are those used in
ref.~\cite{Ciuchini:2002uv} Concerning SUSY parameters, we fix
$m_{\tilde q}=m_{\tilde g}=350$ GeV and consider different
possibilities for the mass insertions.

\begin{figure}[t]
  \begin{center}
    \begin{tabular}{c c}
      \includegraphics[width=0.4\textwidth]{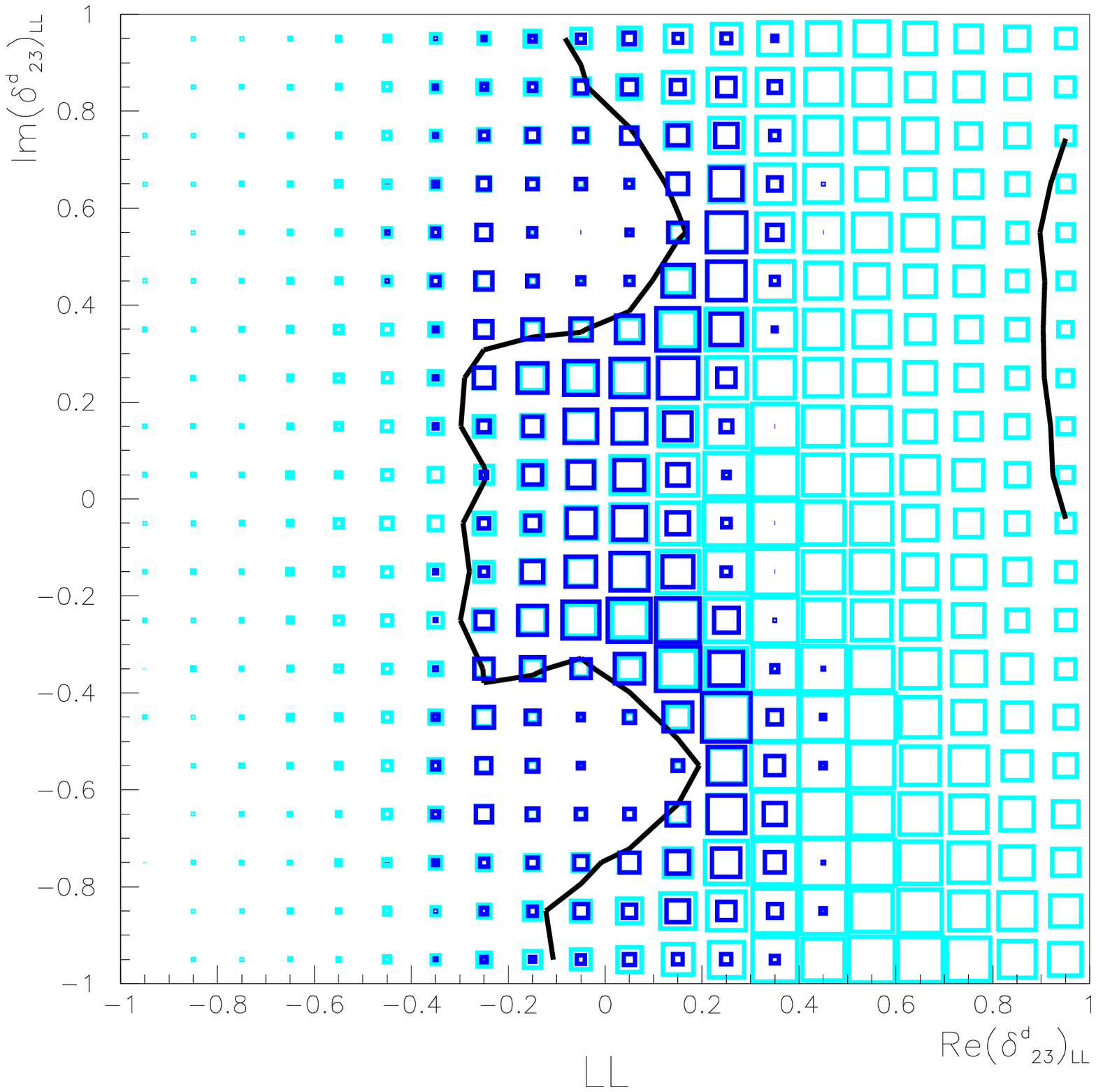} &
      \includegraphics[width=0.4\textwidth]{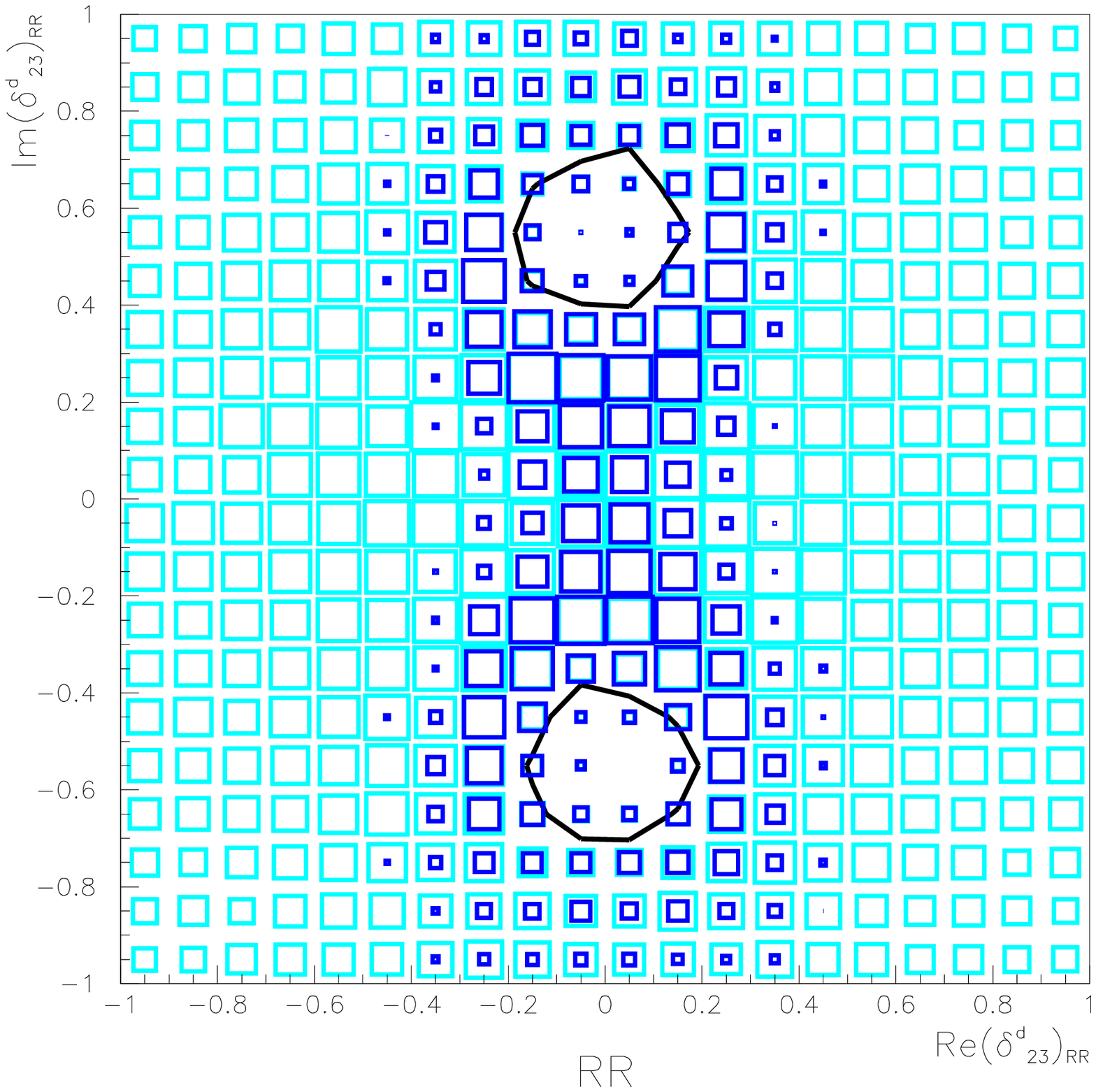} \\
      \includegraphics[width=0.4\textwidth]{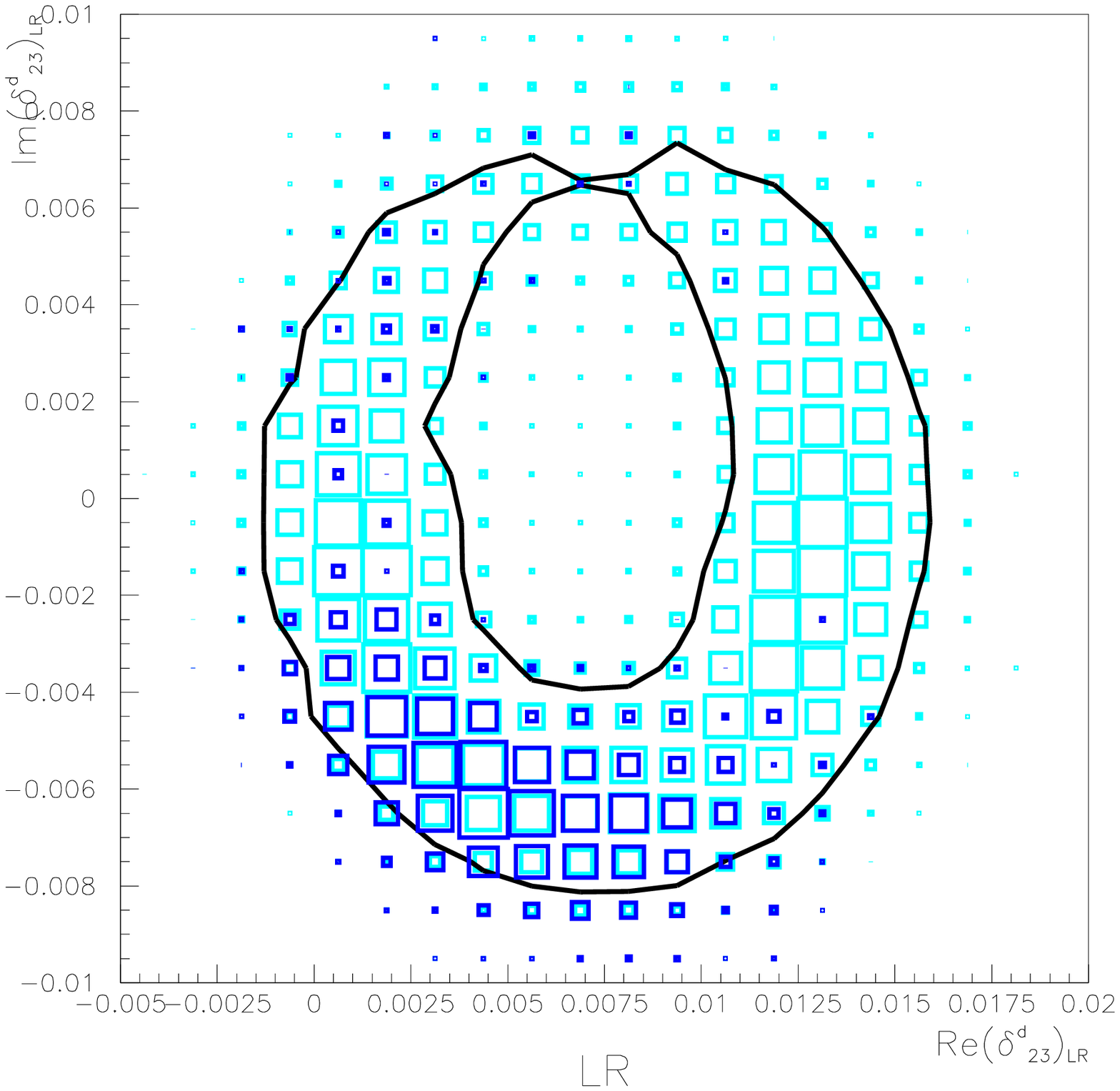} &
      \includegraphics[width=0.4\textwidth]{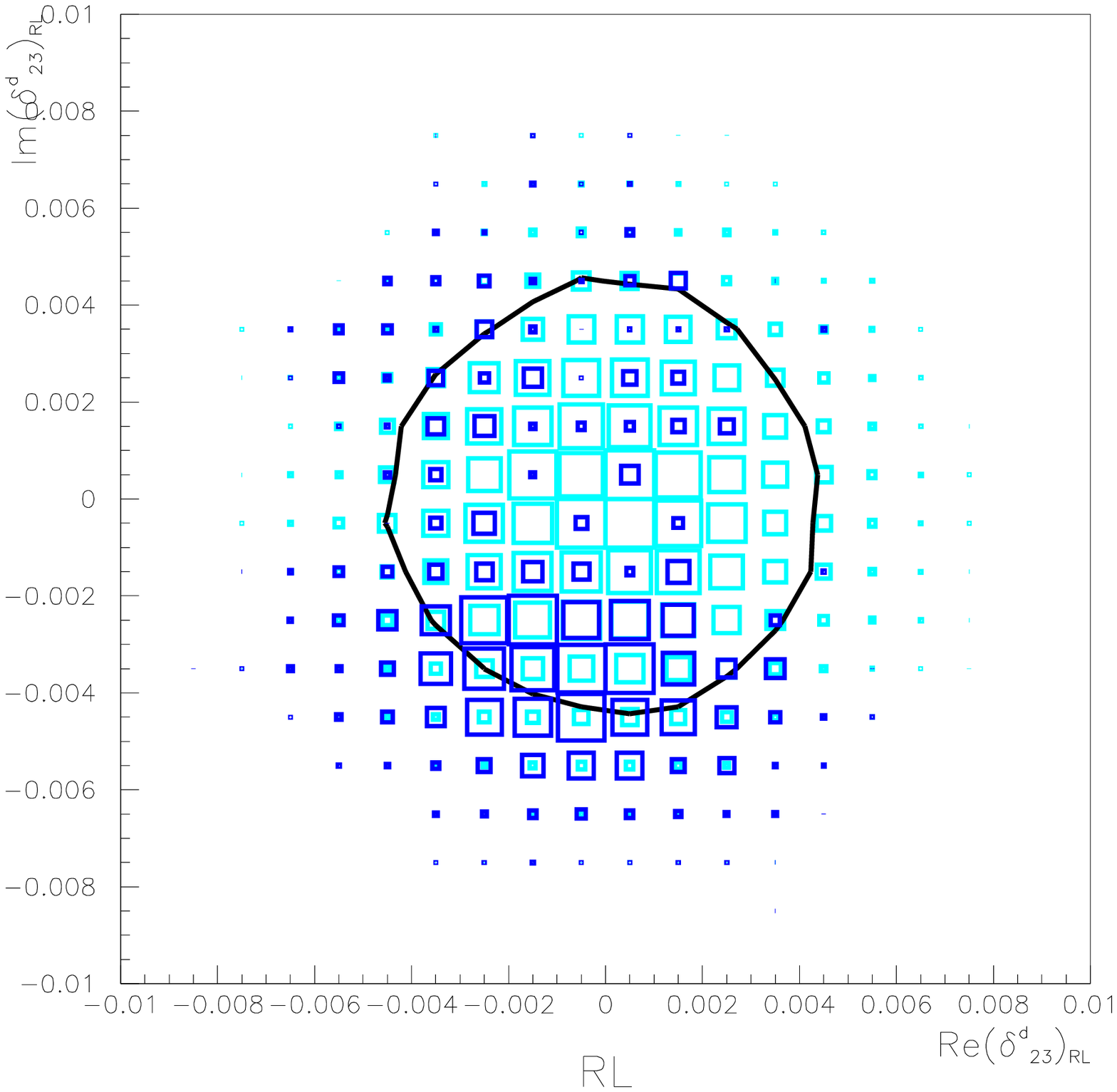} \\ 
    \end{tabular}
  \end{center}
  \caption{Allowed regions in the  
    Re$(\delta^d_{23})_{AB}$--Im$(\delta^d_{23})_{AB}$ space for
    $AB=(LL,RR,LR,RL)$. The black line contains $68 \%$ of the
    weighted events. The darker regions are selected imposing $\Delta
    m_s<20$ ps$^{-1}$ for $LL$ and $RR$ insertions and $S_{\phi K}<0$
    for $LR$ and $RL$ insertions.}
  \label{fig:ranges1}
\end{figure}

In fig.~\ref{fig:ranges1} we display the clustering of events in the
Re$(\delta^d_{23})_{AB}$--Im$(\delta^d_{23})_{AB}$ plane. Here and in the
following plots, larger boxes correspond to larger numbers of weighted
events. The darker regions are selected imposing the further
constraint $\Delta M_s<20$~ps$^{-1}$ for $LL$ and $RR$ insertions and
$S_{\phi K}<0$ for $LR$ and $RL$ insertions.  For helicity conserving
insertions, the constraints are of order $1$. A significant reduction
of the allowed region appears if the cut on $\Delta M_s$ is imposed.
The asymmetry of the $LL$ plot is due to the interference with the SM
contribution. In the helicity flipping cases, constraints are of order
$10^{-2}$. For these values of the parameters, $\Delta M_s$ is
unaffected. We show the effect of requiring $S_{\phi K}<0$: it is
apparent that a nonvanishing Im$\,\delta_{23}^d$ is needed to meet
this condition.

\begin{figure}[t]
  \begin{center}
    \begin{tabular}{c c}
      \includegraphics[width=0.4\textwidth]{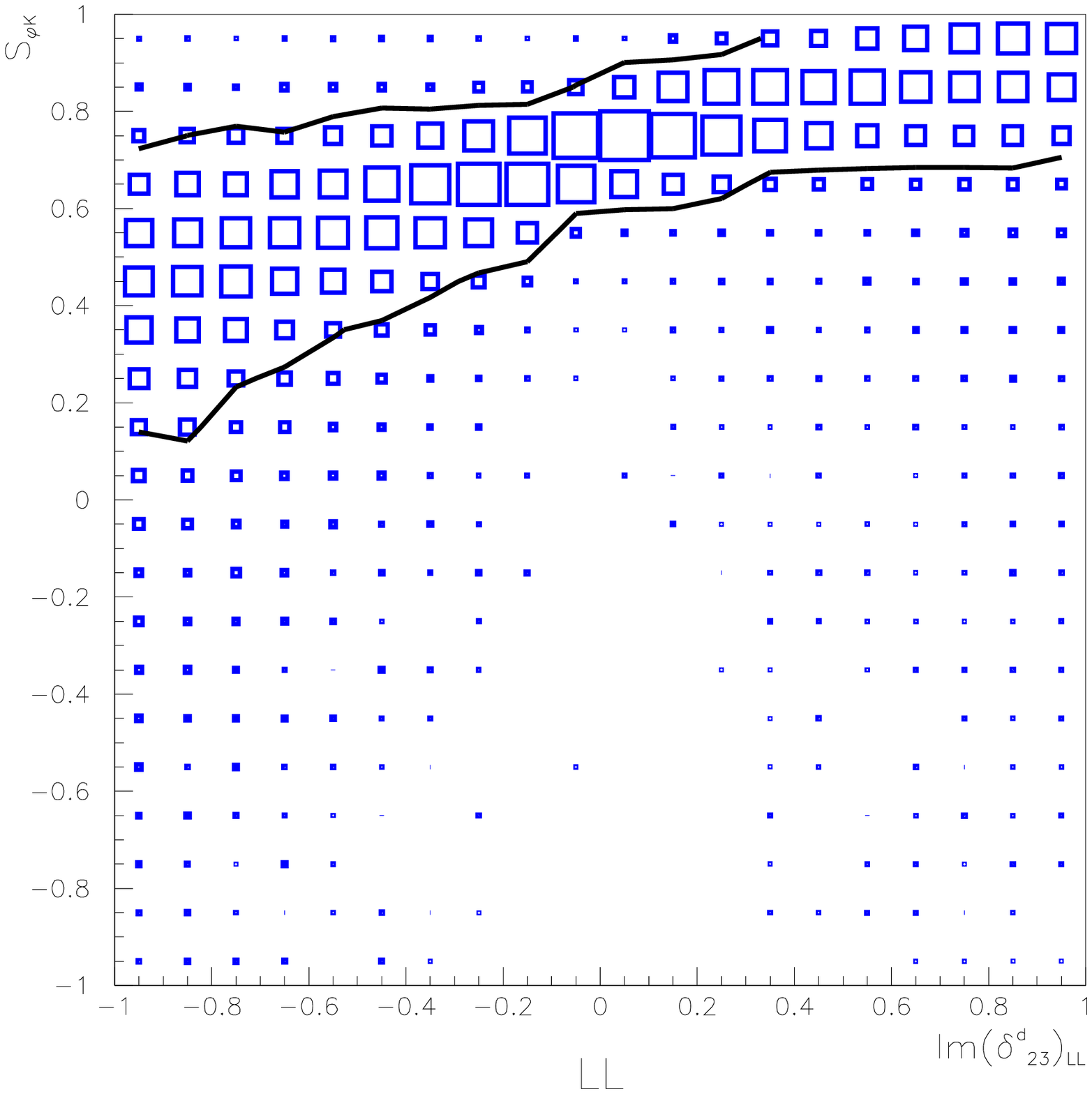} &
      \includegraphics[width=0.4\textwidth]{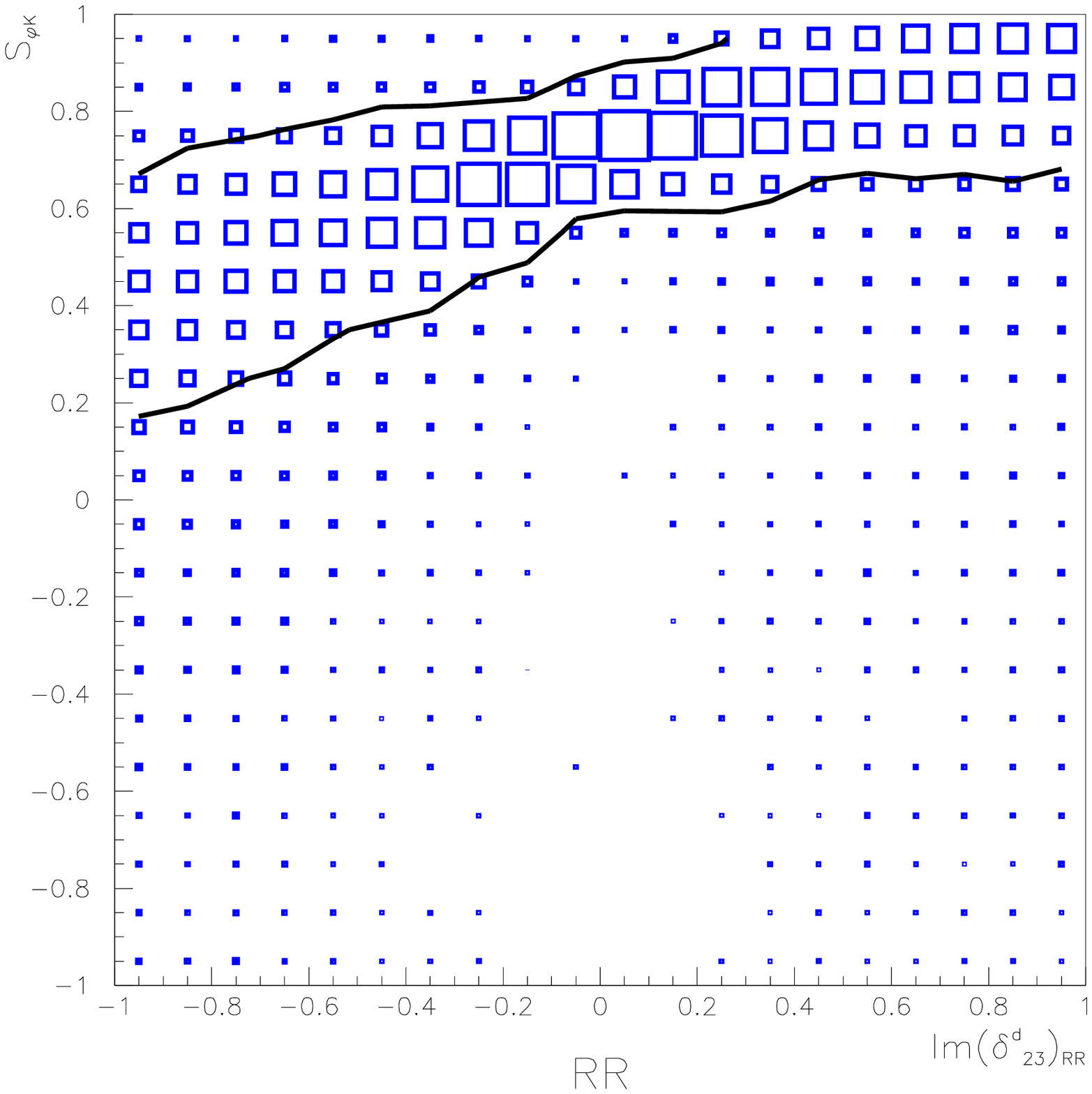} \\
      \includegraphics[width=0.4\textwidth]{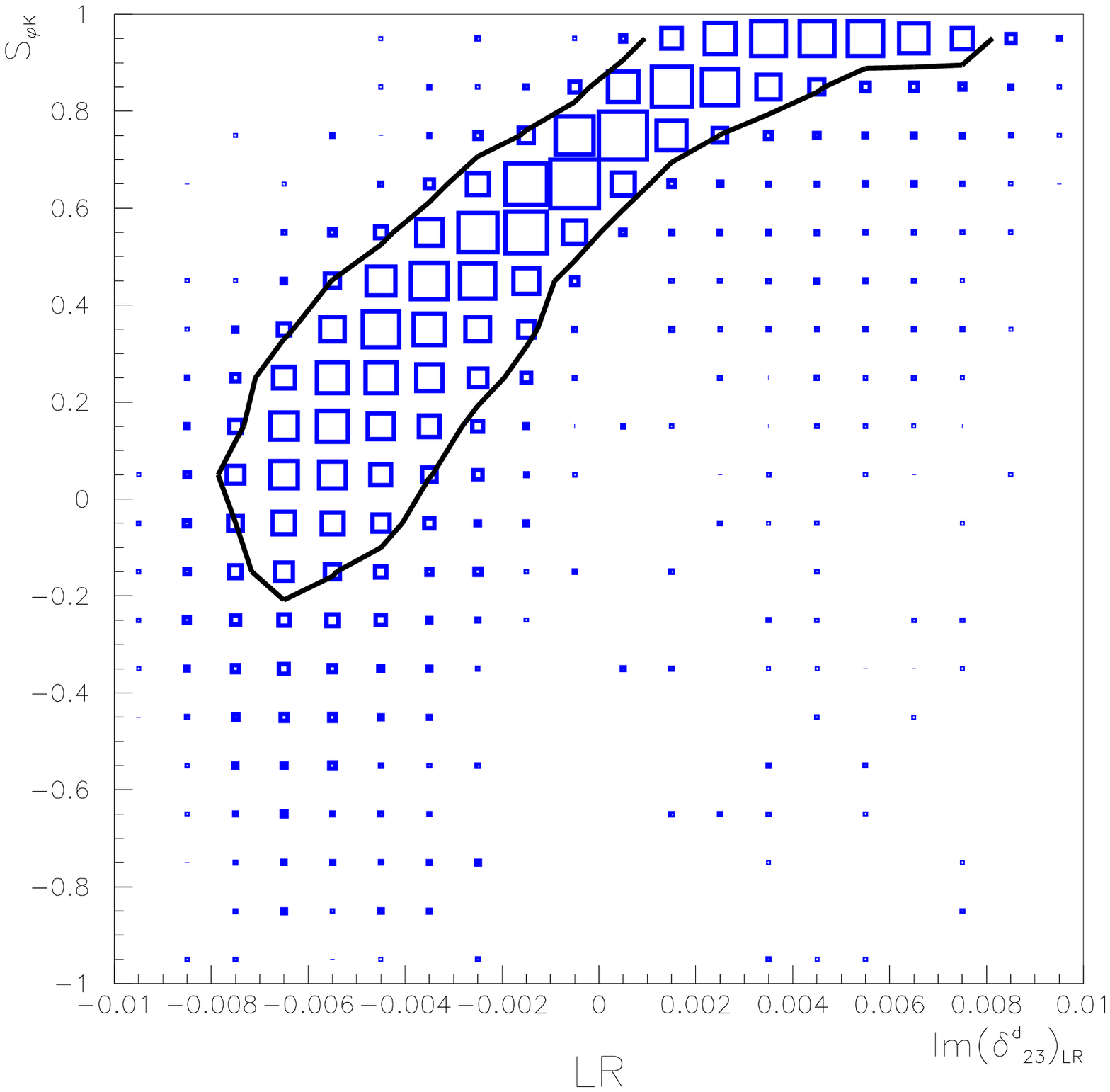} &
      \includegraphics[width=0.4\textwidth]{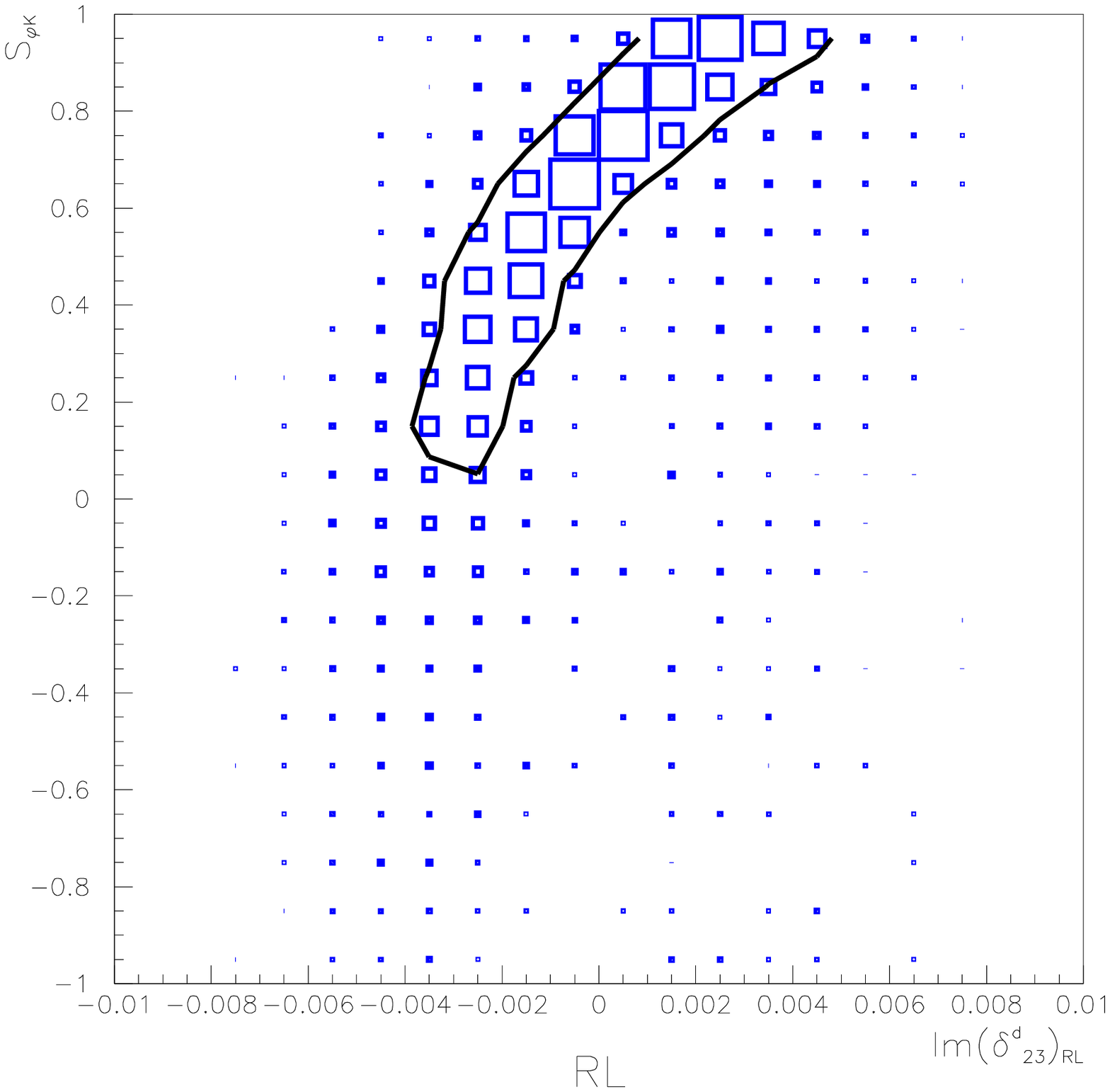} \\ 
    \end{tabular}
  \end{center}
  \caption{Correlations between $S_{\phi K}$ and
      Im$(\delta^d_{23})_{AB}$. The black line 
    contains $68 \%$ of the weighted events.  }
  \label{fig:sinim1}
\end{figure}

In fig.~\ref{fig:sinim1}, we study the correlations of $S_{\phi K}$
with Im$(\delta^d_{23})_{AB}$ for the various SUSY insertions
considered in the present analysis. The reader should keep in mind
that, in all the results reported in fig.~\ref{fig:sinim1}, the
hadronic uncertainties affecting the estimate of $S_{\phi K}$ are not
completely under control. Low values of $S_{\phi K}$ can be more
easily obtained with helicity flipping insertions. A deviation from
the SM value for $S_{\phi K}$ requires a nonvanishing value of
Im$\,(\delta^d_{23})_{AB}$, generating, for those channels in which
the SUSY amplitude can interfere with the SM one, a $A_{CP}(B\to
X_s\gamma)$ at the level of a few percents in the LL case,
and up to the experimental upper bound in the LR case.

\section{Conclusions}
\label{sec:concl}

Two-body nonleptonic $B$ decays are not only a very interesting
theoretical playground to test our understanding of hadronic dynamics,
but also offer a precious window on NP. If any of the present
discrepancies between experimental values and
theoretical predictions will be confirmed in the future, it will be
possible to test NP models against experimental data. As we have
shown, SUSY model can accommodate the Belle result for $\phi K_s$, and
one can find interesting correlation with other observables in $B$ physics.

\end{document}